\documentclass[graybox]{svmult}
\usepackage[fleqn]{amsmath}
\usepackage{amssymb}
\usepackage{mathptmx}       
\usepackage{helvet}         
\usepackage{courier}        
\usepackage{graphicx}        
\usepackage[bottom]{footmisc}

\DeclareMathOperator{\Tr}{Tr}
\DeclareMathOperator{\sgn}{sgn}

\let\Im\relax\DeclareMathOperator{\Im}{Im}
\providecommand{\vev}[1]{\langle#1\rangle}
\providecommand{\inner}[2]{\left(#1,#2\right)}
\providecommand{\abs}[1]{\lvert#1\rvert}
\newcommand{\ddd}[1]{\frac{\delta}{\delta #1}}
\newcommand{\at}{\Big\vert}
\newcommand{\dl}{\partial_\Lambda}
\newcommand{\G}{\mathcal{G}}
\newcommand{\N}{\mathcal{N}}

\newcommand{\NRG}{\text{NRG}}
\def\k{{\vec k}}
\def\up{\uparrow}
\def\down{\downarrow}
\begin{document}
\title*{A gentle introduction to the\\
  functional renormalization group:\\
  the Kondo effect in quantum dots}
\titlerunning{A gentle introduction to the funRG:
  the Kondo effect in quantum dots}
\author{Sabine Andergassen, Tilman Enss, Christoph Karrasch and Volker Meden}
\institute{Sabine Andergassen \at
  Max-Planck-Institut f\"ur Festk\"orperforschung,
  D-70569 Stuttgart, Germany, and \\
  Laboratoire d'Etudes des Propri\'et\'es Electroniques
  des Solides, CNRS, BP 166, 38042 Grenoble, France 
  \and Tilman Enss \at
  INFM--SMC--CNR and Dipt.\ di Fisica,
  Universit\`a di Roma ``La Sapienza'', P.le Aldo Moro 5,\\
  I-00185 Roma, Italy
  \and Christoph Karrasch and Volker Meden \at
  Institut f\"ur Theoretische Physik, Universit\"at G\"ottingen,
  Friedrich-Hund-Platz 1, \\ D-37077 G\"ottingen, Germany}
\maketitle

\abstract{The functional renormalization group provides an efficient
  description of the interplay and competition of correlations on
  different energy scales in interacting Fermi systems.  An exact
  hierarchy of flow equations yields the gradual evolution from a
  microscopic model Hamiltonian to the effective action as a function
  of a continuously decreasing energy cutoff.  Practical
  implementations rely on suitable truncations of the hierarchy, which
  capture nonuniversal properties at higher energy scales in addition
  to the universal low-energy asymptotics.  As a specific example we
  study transport properties through a single-level quantum dot coupled to
  Fermi liquid leads. In particular, we focus on the temperature $T=0$
  gate voltage dependence of the linear conductance.  A comparison
  with exact results shows that the functional renormalization group 
  approach captures the broad resonance plateau as well as the
  emergence of the Kondo scale.  It can be easily extended to more
  complex setups of quantum dots.}


\section{Introduction}
\label{sec:intro}

The coupling of a quantum dot with spin degenerate levels and local
Coulomb interaction (modeled, e.g., by the single-impurity Anderson
model; SIAM) to metallic leads gives rise to Kondo
physics \cite{Hewson}.  At low temperatures and for sufficiently high
barriers the local Coulomb repulsion $U$ leads to a broad resonance
plateau in the linear conductance $G$ of such a setup as a function of
a gate voltage $V_g$ which linearly shifts the level
positions \cite{Tsvelik83,GlazmanRaikh,NgLee,Costi94,Gerland00}. It
replaces the Lorentzian found for noninteracting electrons.  On
resonance the dot is half-filled implying a local spin-$\frac{1}{2}$
degree of freedom responsible for the Kondo effect \cite{Hewson}. For
the SIAM the zero temperature conductance is proportional to the
one-particle spectral weight of the dot at the chemical
potential \cite{MeirWingreen}. The appearance of the plateau in the
conductance is due to the pinning of the Kondo resonance in the
spectral function at the chemical potential for $-U/2\lesssim
V_g\lesssim U/2$ (here $V_g=0$ corresponds to the half-filled dot
case) \cite{Hewson,Gerland00}. Kondo physics in transport through
quantum dots was confirmed experimentally \cite{Goldhaber98,vdWiel},
and theoretically using the Bethe ansatz \cite{Tsvelik83,Gerland00} and
the numerical renormalization-group (NRG)
technique \cite{Wilson75,Krishnamurthy80}.  However, both methods can
hardly be used to study more complex setups.  In particular, the
extension of the NRG to more complex geometries beyond single-level
quantum-dot systems \cite{Costi94,Gerland00,Hofstetter01} is restricted
by the computational complexity which increases sharply with the
number of interacting degrees of freedom.  Alternative methods which
allow for a systematic investigation are therefore required.  We here
propose the functional renormalization group (fRG) approach to study
low-temperature transport properties through mesoscopic systems with
local Coulomb correlations.

A particular challenge in the description of quantum dots is their
distinct behavior on different energy scales, and the appearance of
collective phenomena at new energy scales not manifest in the
underlying microscopic model.  An example of this is the Kondo effect
where the interplay of the localized electron spin on the dot and the
spin of the lead electrons leads to an exponentially (in $U$) small
scale $T_K$. This diversity of scales cannot be captured by
straightforward perturbation theory.  One tool to cope with such
systems is the renormalization group: by treating different energy
scales successively, one can often find an efficient description for
each one.  We will here give an introduction to one particular
variant, the fermionic fRG \cite{SH01}, which is formulated in terms of
an exact hierarchy of coupled flow equations for the vertex functions
as the energy scale is lowered.  The flow starts directly from the
microscopic model, thus including nonuniversal effects on higher
energy scales from the outset, in contrast to effective field theories
capturing only the asymptotic behavior.  As the cutoff scale is
lowered, fluctuations at lower energy scales are successively included
until one finally arrives at the full effective action (the generating
functional of the one-particle irreducible vertex functions) from
which all physical properties can be extracted.  This allows to
control infrared singularities and competing instabilities in an
unbiased way.  Truncations of the flow equation hierarchy and suitable
parametrizations of the frequency and momentum dependence of the
vertex functions lead to powerful new approximation schemes, which are
justified for weak renormalized interactions. A comparison with exact
results shows that the fRG is remarkably accurate even for sizeable
interactions.

The outline of this article is as follows. In Sec.~\ref{sec:frg} we
introduce the fRG formalism and derive the hierarchy of flow
equations.  The implementation of the fRG technique for a quantum dot
modeled by a SIAM is described in Sec.~\ref{sec:siam}. In the
following we present results for the linear conductance including a
comparison with exact solutions and discuss the emergence of the Kondo
scale.  We conclude in Sec.~\ref{sec:outlook} with a summary and
outlook on further applications and extensions of the present work.


\section{The fRG technique}
\label{sec:frg}

To make this article self-contained, this section aims to give a very
short introduction into the fRG \emph{formalism} by deriving the
functional flow equations.  Readers not so much interested in the
formal beauty might jump ahead to the following Sec.\ \ref{sec:siam}
where these flow equations are applied to a specific physical model.
An introduction to the many-body formalism used here can be found,
e.g., in \cite{NegeleOrland}; for the details of the derivation
of the functional flow equation see, e.g., 
\cite{EnssThesis,MetznerSalerno} and references therein.  Exact
functional flow equations were derived for bosonic field theories in
\cite{WH73,Pol84,Wet93} and for fermionic fields in the
one-particle irreducible scheme in \cite{SH01}.

\subsection{Green and vertex functions}
\label{sec:rg:form}
We consider a system of interacting fermions described by Grassmann
variables $\psi$, $\bar\psi$, and an action
\begin{align}
  \label{eq:action}
  S[\psi,\bar\psi] = \inner{\bar\psi}{C^{-1}\psi} - V[\psi,\bar\psi]
\end{align}
with bare propagator
\begin{align}
  C(K) = \frac{1}{i\epsilon-\xi_k}
\end{align}
where the index $K=(\epsilon,\k,\sigma)$ collects the Matsubara
frequency $\epsilon$ and the quantum numbers of a suitable
single-particle basis set, e.g., momentum $\k$ and spin projection
$\sigma$, and $\xi_\k=\epsilon_\k-\mu$ denotes the energy relative to
the chemical potential.  In order to understand the general structure
of the flow equation it is useful to include a Nambu (particle/hole)
index into $K$: then each directed fermion line denotes the
propagation of either a particle or a hole \cite{SH01}.  The inner
product implies a summation over these indices: for our diagonal
propagator, $\inner{\bar\psi}{C^{-1}\psi} = \sum_{K} \bar\psi_{K}
C_{K}^{-1} \psi_K$.  $V[\psi,\bar\psi]$ is an arbitrary many-body
interaction; we will see a specific example for this below in
Eq.~\eqref{eq:siam}.  Connected Green functions can be obtained from
the generating functional \cite{NegeleOrland}
\begin{align}
  \label{eq:G}
  \G[\eta,\bar\eta]
  = -\log \left\{ \N \int D\bar\psi\psi\, e^{S[\psi,\bar\psi]}\,
  e^{-\inner{\bar\psi}{\eta}-\inner{\bar\eta}{\psi}} \right\}
\end{align}
by taking derivatives with respect to the source fields $\eta$:
\begin{align*}
  G_m(K_1',\dots,K_m';K_1,\dots,K_m)
  & = -\vev{\psi_{K_1'}\dots\psi_{K_m'}
    \bar\psi_{K_m}\dots\bar\psi_{K_1}}_{\text{conn}} \\
  & = \frac{\delta^m}{\delta\eta_{K_1}\dots\delta\eta_{K_m}}
  \frac{\delta^m}{\delta\bar\eta_{K_m'}\dots\delta\bar\eta_{K_1'}}
  \G[\eta,\bar\eta]\at_{\eta=\bar\eta=0} \;.
\end{align*}
Equivalently, $\G[\eta,\bar\eta]$ can be expanded in powers of the
source fields with expansion coefficients $G_m(K_1',\dots, K_m)$.  The
normalization factor $\N=\det C$ cancels the noninteracting vacuum
diagrams, such that $\G[0,0]=0$ in the absence of interaction.  As we
explain below, it will be of advantage in our case to describe the
system not by connected Green functions but by the one-particle
irreducible (1PI) vertex functions.  Their generating functional, the
\emph{effective action}, is obtained from $\G$ by Legendre
transformation,
\begin{align}
  \label{eq:gammadef}
  \Gamma[\phi,\bar\phi]
  = \G[\eta,\bar\eta] +\inner{\bar\phi}{\eta} -\inner{\bar\eta}{\phi}.
\end{align}
This functional can again be expanded in powers of the fields $\phi$,
$\bar\phi$ to obtain the vertex functions $\gamma_m(K_1',\dots, K_m)$.
The usual relations between $\G$ and $\Gamma$ hold, i.e.,
$\phi=\delta\G/\delta\bar\eta$, $\bar\phi=\delta\G/\delta\eta$ as well
as $\bar\eta=\delta\Gamma/\delta\phi$,
$\eta=\delta\Gamma/\delta\bar\phi$, and
$\delta^2\Gamma/\delta\phi\delta\bar\phi =
(\delta^2\G/\delta\eta\delta\bar\eta)^{-1}$.

\subsection{Functional flow equations}
We introduce an infrared cutoff into the bare propagator that
suppresses all soft modes, which may be a source of divergences in
perturbation theory.  In bulk systems it is convenient to use a
momentum cutoff, which suppresses momenta close to the Fermi surface.
On the other hand, if translation invariance is spoiled by impurities
or a particular spatial setup as for quantum dots, it is easier to use
a frequency cutoff, excluding all Matsubara frequencies below scale
$\Lambda$ using a step function $\Theta(x)$:
\begin{align}
  \label{eq:Ccutoff}
  C^\Lambda(K) = \frac{\Theta(\abs\epsilon-\Lambda)} {i\epsilon-\xi_k} \,.
\end{align}
This changes the microscopic model to exclude soft modes; in the limit
$\Lambda\to 0$ the original model is recovered.  The easiest way to
understand how the Green functions change with $\Lambda$ is to derive
the flow equation for the cutoff-dependent generating functional
\begin{align}
  \label{eq:Gl}
  \G^\Lambda[\eta,\bar\eta]
  = -\log \left\{ \N^\Lambda \int D\bar\psi\psi\,
    e^{\inner{\bar\psi}{Q^\Lambda\psi}-V[\psi,\bar\psi]}\, 
  e^{-\inner{\bar\psi}{\eta}-\inner{\bar\eta}{\psi}} \right\}
\end{align}
with $(C^\Lambda)^{-1}\equiv Q^\Lambda$, and the normalization factor
$\N^\Lambda=(\det Q^\Lambda)^{-1}$.  Taking the $\Lambda$ derivative
and denoting $\dl Q^\Lambda=\dot Q^\Lambda$,
\begin{align}
  \dl\G^\Lambda
  & = -\dl \log \N^\Lambda
  - \frac{1}{e^{-\G^\Lambda}} \, \int D\bar\psi\psi\, 
  \inner{\bar\psi}{\dot Q^\Lambda\psi} \,
  e^{\inner{\bar\psi}{Q^\Lambda\psi}-V[\psi,\bar\psi]
    -\inner{\bar\psi}{\eta}-\inner{\bar\eta}{\psi}} \notag \\
  & = -\dl \log \N^\Lambda + e^{\G^\Lambda}
  \left( \ddd{\eta}, \dot Q^\Lambda \ddd{\bar\eta} \right)
  e^{-\G^\Lambda} \notag \\
  \label{eq:gflow}
  & = \Tr(\dot Q^\Lambda\, C^\Lambda)
  -\Tr\left(\dot Q^\Lambda\,
    \frac{\delta^2 \G^\Lambda}
    {\delta\eta\, \delta\bar\eta}\right)
  +\inner{\frac{\delta\G^\Lambda}{\delta\eta}} {\dot
    Q^\Lambda \frac{\delta\G^\Lambda}{\delta\bar\eta}}
\end{align}
where the first term, $-\dl \log \N^\Lambda = \Tr(\dot Q^\Lambda
C^\Lambda)$, comes from the derivative of the normalization factor,
and the trace denotes a sum over all $K$.

In the present application to quantum dots, the one-particle potential
is strongly renormalized, and it is important to include the feedback
of this renormalization nonperturbatively in the flow equations.  This
is most easily achieved in the one-particle irreducible (1PI) scheme,
where the one-particle renormalizations are included in the internal
propagators.  Hence, we perform a Legendre transform which now also
depends on $\Lambda$,
\begin{align}
  \Gamma^\Lambda[\phi,\bar\phi]
  = \G^\Lambda[\eta^\Lambda,\bar\eta^\Lambda]
  + \inner{\bar\phi}{\eta^\Lambda} - \inner{\bar\eta^\Lambda}{\phi},
\end{align}
and we find $\dl \Gamma^\Lambda = d\G^\Lambda/d\Lambda +
\inner{\bar\phi}{\dl\eta^\Lambda} - \inner{\dl\bar\eta^\Lambda}{\phi}
= \dl \G^\Lambda$ after taking into account the derivatives also of
the $\eta^\Lambda$ fields.  The fundamental variables of
$\Gamma^\Lambda$ are the $\phi$ fields and the $\eta$ acquire a
$\Lambda$ dependence via the relation between $\phi$ and
$\eta$ \cite{NegeleOrland}. By Legendre transform of the flow
Eq.~\eqref{eq:gflow}, we obtain the 1PI flow equation
\begin{align}
  \label{eq:gammaflow}
  \dl \Gamma^\Lambda[\phi,\bar\phi]
  & = \dl \G^\Lambda
  = \Tr \dot Q^\Lambda \Bigl[ C^\Lambda - \Bigl(\frac{\delta^2
    \Gamma^\Lambda[\phi,\bar\phi]}{\delta\phi\,\delta\bar\phi}
  \Bigr)^{-1} \Bigr] + \inner{\bar\phi}{\dot Q^\Lambda\phi} \,.
\end{align}
The inversion of a functional of Grassmann variables with a nonzero
complex part is well defined as a geometric series, which involves
only products of Grassmann variables.  Therefore, we write the Hessian
of $\Gamma^\Lambda$ as
\begin{align}
  \frac{\partial^2\Gamma^\Lambda}{\partial\phi_K\partial\bar\phi_{K'}}
  & = (G^\Lambda)_{K,K'}^{-1} +
  \tilde\Gamma_{K,K'}^\Lambda[\phi,\bar\phi] \,,
\end{align}
where the inverse full propagator $(G^\Lambda)^{-1} = \gamma_1^\Lambda
= Q^\Lambda - \Sigma^\Lambda$ according to the Dyson equation, and
$\tilde\Gamma^\Lambda$ depends at least quadratically on the $\phi$,
$\bar\phi$ fields.  This decomposition allows us to write the inverse
as a geometric series,
\begin{align}
  \left( \frac{\partial^2\Gamma^\Lambda}
    {\partial\phi_K\partial\bar\phi_{K'}} \right)^{-1} 
  & = (1+G^\Lambda \tilde\Gamma^\Lambda)^{-1} G^\Lambda
  = \Bigl( 1 - G^\Lambda\, \tilde\Gamma^\Lambda 
    + [G^\Lambda\, \tilde\Gamma^\Lambda]^2 - \dotsb
  \Bigr) G^\Lambda \,,\notag
\end{align}
which we insert into the flow Eq.~\eqref{eq:gammaflow},
\begin{align}
  \label{eq:gammaflow2}
  \dl \Gamma^\Lambda = \Tr(\dot Q^\Lambda [C^\Lambda-G^\Lambda])
  +\inner{\bar\phi}{\dot Q^\Lambda\phi}
  +\Tr(S^\Lambda [\tilde\Gamma^\Lambda - \tilde\Gamma^\Lambda\,
  G^\Lambda\, \tilde\Gamma^\Lambda\, + \dotsb]) \,.
\end{align}
The first term describes the flow of the zero-point function (grand
canonical potential), while the second term expresses the change of
the bare propagator with $\Lambda$.  The hierarchy of flow equations
is encoded in the third term which represents a loop with an arbitrary
number of vertices $\tilde\Gamma^\Lambda$, which contribute at least
two external legs each, connected by full propagators $G^\Lambda$ and
one single-scale propagator $S^\Lambda$ which has support only at
frequency $\Lambda$,
\begin{align}
  \label{eq:singlescale}
  S^\Lambda = G^\Lambda \dot Q^\Lambda G^\Lambda \,.
\end{align}
Given the initial conditions at $\Lambda=\infty$, the flow equations
determine $\Gamma^\Lambda$ uniquely for all $\Lambda<\infty$.  The
flow of $\Gamma^\Lambda$ thus interpolates between the bare action
$\Gamma^{\Lambda=\infty}=S$ and the full solution of the problem,
$\Gamma^{\Lambda=0}=\Gamma$.

\subsection{Flow equation hierarchy}
\begin{figure}[htbp]
  \centering
  \includegraphics[width=\textwidth]{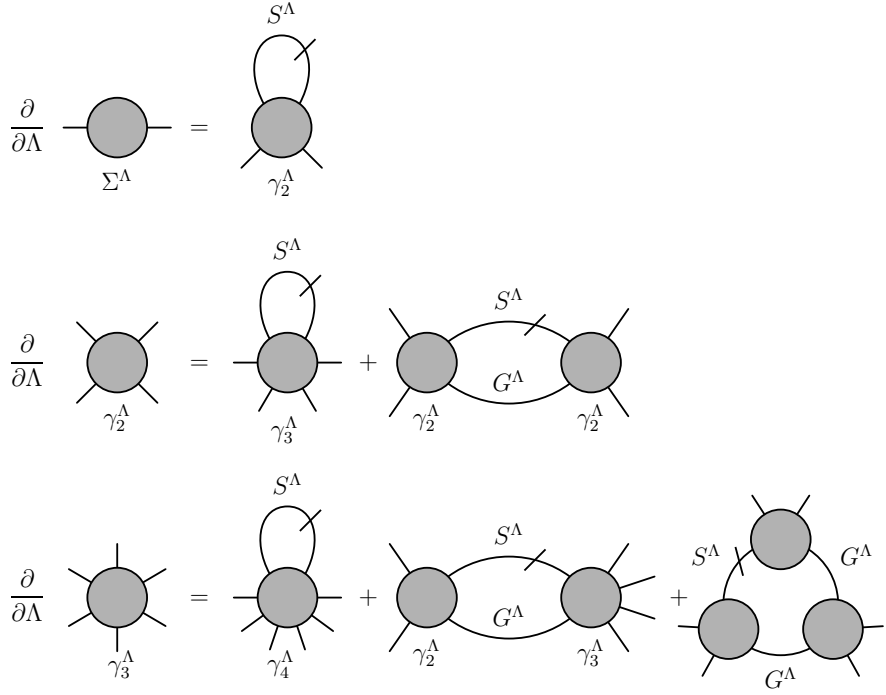}
  \caption{Diagrammatic representation of the flow equations for the
    self-energy $\Sigma^\Lambda$, the two-particle vertex
    $\gamma_2^\Lambda$ and the three-particle vertex
    $\gamma_3^\Lambda$ in the 1PI formulation of the fRG.}
  \label{fig:flow}
\end{figure}

Expanding the functional flow equation for $\Gamma^\Lambda$ in powers
of the source fields $\phi$, $\bar\phi$, we obtain an infinite
hierarchy of coupled flow equations for the $m$-particle vertex
functions $\gamma_m^\Lambda$.  This hierarchy can be represented
diagrammatically and the first three levels are shown in
Fig.~\ref{fig:flow}, where each line represents the propagation of
either a particle or a hole.  For instance, the last diagram in the
second line includes both a particle-hole and a particle-particle
bubble.  For our application to the SIAM it is convenient to
distinguish these\footnote{A derivation of the flow equation using
  separate particle and hole propagators from the outset can be found,
  e.g., in \cite{HMPS04}.}, so we denote by
$\gamma_2(1',2';1,2)$ the two-particle vertex with incoming electrons
$1$, $2$ and outgoing electrons $1'$, $2'$.

The first line of Fig.~\ref{fig:flow} represents the flow equation for
the self-energy $\Sigma^\Lambda$,
\begin{align}
  \label{eq:Sigmaflow}
  \frac{\partial}{\partial\Lambda} \Sigma^\Lambda(1',1)
  & = -T \sum_{2,2'} e^{i\epsilon_2 0^+} \,
  S^\Lambda(2,2') \, \gamma_2^\Lambda(1',2';1,2) \,,
\end{align}
where the labels $1,1',\dots$ are a shorthand notation for
$K_1,K_{1'}$ etc., and the summation includes Matsubara frequencies.
The two-particle vertex $\gamma_2^\Lambda$ entering
Eq.~\eqref{eq:Sigmaflow} is determined by the second line in
Fig.~\ref{fig:flow}, which in turn depends on the three-particle
vertex $\gamma_3^\Lambda$, etc.  This infinite system of coupled
differential equations can usually not be integrated analytically;
therefore one has to resort to numerical computations which require to
truncate the hierarchy by neglecting the flow of the higher vertices.
This may be justified perturbatively for weak interactions since the
higher vertices not present in the bare interaction are generated only
at higher orders in the effective interaction.  As a first step, we
consider only the flow of the self-energy with all higher vertices
remaining at their initial conditions (the bare interaction $V$): in
our application, this already produces qualitatively the correct
result.  As a second step, we also include the flow of
$\gamma_2^\Lambda$ but neglect $\gamma_3^\Lambda$, which brings us
even quantitatively within a few percent of the known results:
\begin{align}
  \label{eq:Vflow}
  & \frac{\partial}{\partial\Lambda} \gamma_{2}^{\Lambda}(1',2';1,2)
  = T \, \sum_{3,3'} \, \sum_{4,4'}
  G^{\Lambda}_{3,3'} \, S^{\Lambda}_{4,4'} \Big[ 
  \gamma_{2}^{\Lambda}(1',2';3,4) \,
  \gamma_{2}^{\Lambda}(3',4';1,2) \\ 
  & \;
  - \gamma_{2}^{\Lambda}(1',4';1,3) \,
  \gamma_{2}^{\Lambda}(3',2';4,2) 
  + \gamma_{2}^{\Lambda}(2',4';1,3) \, 
  \gamma_{2}^{\Lambda}(3',1';4,2) 
  + (3\leftrightarrow 4,3'\leftrightarrow 4')
  \Big] \,, \notag
\end{align}
with the first term representing the particle-particle channel and the
other two the particle-hole channels, respectively.  The last bracket
means that the two terms in the second line have to be repeated with
the changes of variables as indicated.

There is one technical detail which allows to simplify these equations
significantly: the sharp frequency cutoff in Eq.~\eqref{eq:Ccutoff}
has the advantage that the frequency integrals on the right-hand side
of Eqs.~\eqref{eq:Sigmaflow} and \eqref{eq:Vflow} can be performed
analytically.  The propagators contain both
$\Theta(\abs\epsilon-\Lambda)$ and
$\delta(\abs\epsilon-\Lambda)=-\dl\Theta(\abs\epsilon-\Lambda)$
functions, which at first look ambiguous because the step function has
a jump exactly where the $\delta$ function has support.  However, if
one smoothes the step slightly and makes a change of variables from
$\Lambda$ to $t=\Theta(\abs\epsilon-\Lambda)$ and
$dt=-\delta(\abs\epsilon-\Lambda)d\Lambda$ in the vicinity of
$\abs\epsilon=\Lambda$, the integration over the step becomes
well-defined and is implemented by the
substitution \cite{Mor94,EnssThesis}
\begin{align}
  \label{eq:morris}
  \delta(x-\Lambda) f\bigl(\Theta(x-\Lambda)\bigr) \to \delta(x-\Lambda)
  \int_0^1 f(t) dt \,.
\end{align}
Using this substitution, the product of full and single-scale
propagators around the loop is replaced by a delta
function\footnote{There remain additional frequency constraints if not
  all propagators are at the same frequency.}
$\delta(\abs\epsilon-\Lambda)$ times smooth propagators of the form
\begin{align}
  \label{eq:Gtilde}
  \tilde G^\Lambda & = [G_0^{-1}-\Sigma^\Lambda]^{-1} \,.
\end{align}
Note that the explicit $\Lambda$ dependence remains only in the
self-energy, such that the resulting propagator (and hence the flow
equation) is smooth in both $\Lambda$ and $\epsilon$ and can be easily
integrated numerically.


\section{Application to quantum dots}
\label{sec:siam}
In order to give a pedagogical example how the functional flow
equations can be used to solve an interesting physical problem, we
choose as a toy model the SIAM \cite{GlazmanRaikh,NgLee}, which is used
to study, e.g., transport through a quantum dot.  After integrating
out the noninteracting lead degrees of freedom (see below) this model
has zero space dimensions, hence spin is the only quantum number, and
the resulting flow equations are particularly simple and can even be
integrated analytically if the flow of the two-particle vertex is
neglected.  Nevertheless, the physics is nontrivial, and the
exponentially small scale (the Kondo scale) we obtain is seen neither
in perturbation theory in $U$ (which in the present example is free of
infrared divergencies) nor in self-consistent Hartree-Fock
calculations.

\subsection{The single-impurity Anderson model}

\begin{figure}[htbp]
  \sidecaption
  \includegraphics[height=2.5cm]{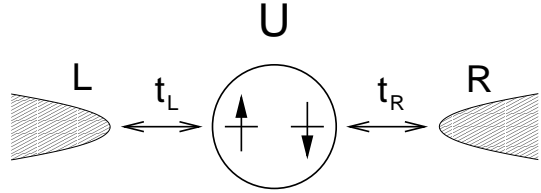}
  \caption{A quantum dot connected to two leads (reservoirs).}
  \label{fig:skizze}
\end{figure}

Our model consists of a single site with local Coulomb repulsion
$U\geq 0$, which is connected to two leads $l=L,R$ via tunnel barriers
$t_{L,R}$ (see Fig.~\ref{fig:skizze}):
\begin{align}
  \label{eq:siam}
  H & = U (n_\up-\tfrac 12) (n_\down-\tfrac 12)
  + \sum_\sigma \epsilon_\sigma d_\sigma^\dagger d_\sigma^{\phantom\dagger}
  - \sum_{\sigma,l} t_l (d_\sigma^\dagger 
c_{0,l}^{\phantom\dagger} + \mbox{H.c.}) \;,
\end{align}
where $d_\sigma^\dagger$ is the creation operator for an electron with
spin $\sigma$ on the dot, $n_\sigma=d_\sigma^\dagger d_\sigma$ is the
spin-$\sigma$ number operator, and $c_l^{\dagger}$ denotes the
creation operator at the end of the semi-infinite lead $l$.  The leads
may be modeled by tight-binding chains, $H_l=-t\sum_{m=0}^\infty
(c^\dagger_{m,l} c_{m+1,l}^{\phantom\dagger} + h.c.)$ with hopping
amplitude $t$.  The hybridization of the dot with the leads broadens
the levels on the dot by $\Gamma_l(\epsilon) = \pi t_l^2
\rho_l(\epsilon)$, where $\rho_l(\epsilon)$ is the local density of
states at the end of lead $l$ which we henceforth assume to be
constant (wide-band limit).  A local magnetic field $h$ at the dot
site splits the spin-up and spin-down energy levels around the gate
voltage $V_g$ ($\sigma_{\up\down}=\pm 1$),
\begin{align}
  \epsilon_{\up\down} & = V_g + \sigma h/2 \;.
\end{align}
Since the leads are noninteracting, we can integrate out the degrees
of freedom of the lead electrons in the path integral and thereby
obtain a hybridization contribution $\Gamma = \Gamma_L + \Gamma_R$ to
the bare Green function of the dot which depends only on the sign of
the Matsubara frequency \cite{Hewson},
\begin{align}
  \label{eq:barepropsiam}
  G_{0,\sigma}(i\epsilon) & = \frac{1}{i\epsilon-(V_g+\sigma
    h/2)+i\Gamma\sgn(\epsilon)} \,.
\end{align}
By solving the interacting many-body problem we obtain a
frequency-dependent self-energy $\Sigma_\sigma(i\epsilon)$ on the dot,
and the full dot propagator is given by the Dyson equation,
\begin{align}
  \label{eq:fullprop}
  G_\sigma(i\epsilon) & =
  [G_{0,\sigma}(i\epsilon)^{-1}-\Sigma_\sigma(i\epsilon)]^{-1} \;.
\end{align}
This self-energy could be computed by perturbation theory in the
strength of the Coulomb interaction, e.g., in a Hartree-Fock
calculation, but it turns out that this does not capture the physical
effect that is observed in experiments and that we wish to describe.
Instead, we will now show how the fRG can be used to compute with very
little effort an approximation for the self-energy that leads to the
correct physical properties of the $T=0$ conductance.

In general the linear response conductance is given by the
current-current correlation function (Kubo formula).  At zero
temperature, zero bias voltage, and for a single interacting site the
exact conductance has the simple form \cite{MeirWingreen}
\begin{align}
  \label{eq:cond}
  G(V_g) & = G_\up(V_g) + G_\down(V_g) \\
  G_\sigma(V_g) & = \frac{e^2}{h} \, \pi\Gamma\rho_\sigma(0) \notag
\end{align}
in terms of the dot spectral function continued analytically to real
frequencies, 
\begin{align}
  \label{eq:rho}
  \rho_\sigma(\epsilon) & = -\frac{1}{\pi} \Im G_\sigma(\epsilon+i0^+) \,.
\end{align}

\subsection{Flow equation for the self-energy}
In order to implement the fRG, following Eq.~\eqref{eq:Ccutoff}
we introduce a cutoff in Matsubara frequency into the bare propagator
on the dot \eqref{eq:barepropsiam},
\begin{align}
  G_{0,\sigma}^\Lambda(i\epsilon)
  & = \frac{\Theta(\abs\epsilon-\Lambda)}
  {i\epsilon-(V_g+\sigma h/2)+i\Gamma\sgn(\epsilon)} \;.
\end{align}
This is an infrared cutoff which sets the propagator to zero for
frequencies smaller than $\Lambda$ (preventing these modes from being
excited), but leaves the high-energy modes unchanged (propagating).
As the cutoff scale $\Lambda$ is gradually lowered, more and more
low-energy degrees of freedom are included, until finally the original
model is recovered for $\Lambda\to 0$.  Changing the cutoff scale
leads to the infinite hierarchy of flow equations for the vertex
functions shown in Fig.~\ref{fig:flow}.  We use the flow equation for
the self-energy \eqref{eq:Sigmaflow} with the single-scale propagator
substituted by \eqref{eq:Gtilde}.  The two-particle vertex
$\gamma_2^\Lambda$ is in general a complicated function of three
independent frequencies (justifying the name \emph{functional}
renormalization group) that evolves during the flow by the second line
of Fig.~\ref{fig:flow}.  As a first approximation we neglect the flow
of the two-particle vertex and all higher vertex functions.  This can
be justified perturbatively if the bare coupling $U$ is small, as all
the terms generated during the flow are of higher order in $U$.  As a
consequence, the two-particle vertex
\begin{align}
  \label{eq:Uparam}
  \gamma_2^\Lambda(i\epsilon\sigma,i\epsilon'\bar\sigma;
  i\epsilon\sigma,i\epsilon'\bar\sigma) = U^\Lambda
\end{align}
remains for all $\Lambda$ at its initial condition, $U^\Lambda\equiv
U^{\Lambda=\infty}=U$ which is the bare Coulomb interaction in the
Hamiltonian \eqref{eq:siam}.  We denote $\bar\sigma=-\sigma$.  As
$\gamma_2^\Lambda$ does not depend on frequency in our approximation,
also the self-energy does not acquire a frequency dependence during
the flow.  The flow equation for the effective level position
$V_\sigma^\Lambda = V_g + \sigma h/2 + \Sigma_\sigma^\Lambda$ is
\begin{align}
\label{eq:diffkondo}
  \dl V_\sigma^\Lambda & = -\frac{U^\Lambda}{2\pi}
  \int d\epsilon\, \delta(\abs\epsilon-\Lambda) \,
  \tilde G_{\bar\sigma}^\Lambda(i\epsilon)
  = -\frac{U^\Lambda}{2\pi}
  \sum_{\epsilon=\pm\Lambda}
  \frac{1}{i\epsilon-V_{\bar\sigma}^\Lambda+i\Gamma\sgn\epsilon}
  \notag \\
  & = \frac{U^\Lambda V_{\bar\sigma}^\Lambda/\pi}
  {(\Lambda+\Gamma)^2+(V_{\bar\sigma}^\Lambda)^2}
\end{align}
with initial condition $V_\sigma^{\Lambda=\infty} = V_g+\sigma
h/2$ \cite{KEM06}.  At the end of the flow, the renormalized
potential $V_\sigma = V_\sigma^{\Lambda=0}$ determines the dot
spectral function \eqref{eq:rho},
\begin{align}
  \rho_\sigma(\epsilon)
  & = \frac{1}{\pi} \,
  \frac{\Gamma}{(\epsilon-V_\sigma)^2+\Gamma^2} \,,
\end{align}
which is a Lorentzian of full width $2\Gamma$ and height $1/\pi\Gamma$
centered around $V_\sigma$.  Although the true spectral function has a
very different form with a very sharp Kondo peak at
$\epsilon=V_\sigma$, this difference is not seen in the $T=0$
conductance \eqref{eq:cond}: it probes only the value of the spectral
function \emph{at} the chemical potential $\epsilon=0$,
\begin{align}
  G_\sigma(V_g) & = \frac{e^2}{h} \, 
  \frac{\Gamma^2}{V_\sigma^2 + \Gamma^2} \,.
\end{align}
In the noninteracting case, $V_\sigma=V_g+\sigma h/2$ and the
conductance is a sum of two Lorentzians of the applied gate voltage.
If interaction is switched on, this changes drastically.

\begin{figure}[htbp]
  \centering
  \includegraphics[height=10cm,angle=-90]{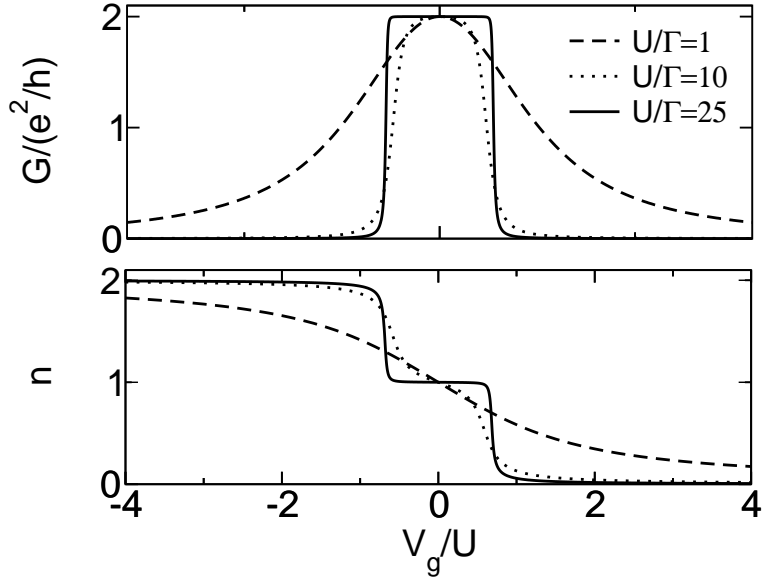}
  \caption{\emph{Upper panel:} conductance as 
    a function of gate voltage for different values of $U/\Gamma$ and
    $h=0$.  \emph{Lower panel:} average number of electrons
    on the dot.}
  \label{fig:fig3}
\end{figure}

In Fig.~\ref{fig:fig3} the conductance $G$ as a function of gate
voltage $V_g$ for the SIAM is shown for different values of $U/\Gamma$
in the upper panel, together with the occupation of the dot in the
lower, both for the case without magnetic field, implying
$V_\up^\Lambda = V_\down^\Lambda =V^\Lambda$ and $G_\up = G_\down$.
For $\Gamma \ll U$ the resonance exhibits a plateau \cite{Gerland00}.
In this region the occupation is close to $1$ while it sharply
rises/drops to $2$/$0$ to the left/right of the plateau.  Also for
asymmetric barriers we reproduce the exact resonance height $4\Gamma_L
\Gamma_R/(\Gamma_L+\Gamma_R)^2\,(2e^2/h)$ \cite{Hewson,Gerland00}. Here
we focus on strong couplings $U/\Gamma \gg 1$.

\begin{figure}[htbp]
  \centering
  \includegraphics[height=9.5cm,angle=-90,clip]{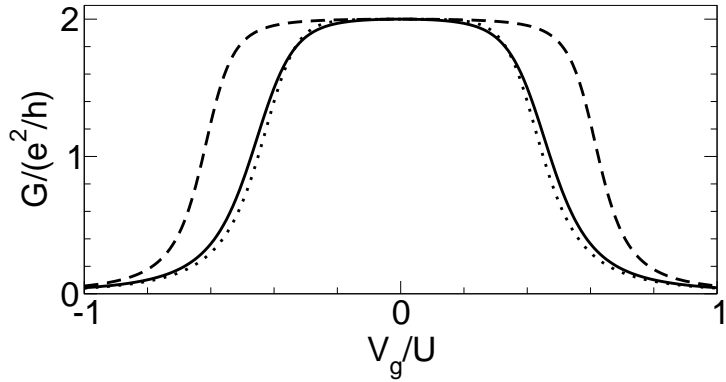}
  \caption{Gate voltage dependence of the conductance at $U/\Gamma=4 \pi$ and
    $h=0$. \emph{Solid line:} exact Bethe ansatz solution from
    \cite{Gerland00}. \emph{Dashed line:} fRG approximation
    without flow of the interaction. \emph{Dotted line:} fRG
    approximation with flow of vertex.}
  \label{fig:fig4}
\end{figure}

The solution of the differential Eq.~(\ref{eq:diffkondo}) at
$\Lambda=0$ is obtained in implicit form
\begin{align}
  \label{eq:kondosol}
  \frac{vJ_1(v)-\gamma J_0(v)}{vY_1(v)-\gamma Y_0(v)}=
  \frac{J_0(v_g)}{Y_0(v_g)}\; ,
\end{align}
with $v = V\pi/U$, $v_g = V_g\pi/U$, $\gamma = \Gamma \pi/U$, and
Bessel functions $J_n$, $Y_n$. For $\abs{V_g} < V_c$ this equation has
a solution with small $\abs{V}$, where $v_c = V_c \pi /U$ is the first
zero of $J_0$ corresponding to $V_c \simeq 0.77\, U$.  For $U\gg
\Gamma$ the crossover to a solution with $\abs{V}$ being of order $U$
(for $\abs{V_g}> V_c$) is fairly sharp. Expanding both sides of
Eq.~\eqref{eq:kondosol} for small $\abs{v}$ and $\abs{v_g}$ gives
\begin{align}
 \label{eq:exp}
  V = V_g \exp \Big( -\frac{U}{\pi \Gamma}\Big) \; .
\end{align}
The consequent exponential pinning of the spectral weight in
Eq.~\eqref{eq:rho} at the chemical potential for small $\abs{V_g}$ and
the sharp crossover to a $V$ of order $U$ when $\abs{V_g} > V_c$ leads
to the observed resonance line shape represented by the dashed line in
Fig.~\ref{fig:fig4}.  For $U\gg \Gamma$ the width of the plateau is
$2V_c \simeq 1.5\, U$, which is larger than the width $U$ found with
the Bethe ansatz \cite{Tsvelik83,Gerland00} corresponding to the solid
line in Fig.~\ref{fig:fig4}. Our approximation furthermore slightly
overestimates the sharpness of the box-shaped resonance.  The
inclusion of the renormalization of the two-particle vertex improves
the quantitative accuracy of the results considerably (see below),
while the pinning of the spectral function and the subsequent
resonance plateau is captured already at the first order of the
flow-equation hierarchy, even though the spectral function neither
shows the narrow Kondo resonance nor the Hubbard bands.

\subsection{Flow equation for the fermion interaction}

With the simple approximation above, we have obtained an exponentially
small scale for the local potential from a renormalization group flow
equation.  This yields a plateau in the conductance that qualitatively
agrees already well with the exact Bethe ansatz solution.  We will now
demonstrate how one can go beyond the simplest approximations in the
renormalization group flow equations to improve (systematically for
small $U$) the result and the agreement with the known solutions.  Any
improvement of the flow equation has to come from a more detailed
parametrization of the two-particle vertex.  The next step is to
include the flow of the two-particle vertex $\gamma_2^\Lambda$ from
the second line in Fig.~\ref{fig:flow}, but still neglect the
three-particle vertex $\gamma_3^\Lambda$ and the frequency dependence
of $\gamma_2^\Lambda$.\footnote{Including the frequency dependence of
  $\gamma_2^\Lambda$ is an ambitious project that is beyond the scope
  of this introductory article \cite{HMPS04}.} Using the vertex flow
Eq.~\eqref{eq:Vflow} with a sharp cutoff and the parametrization
\eqref{eq:Uparam}, we obtain \cite{KEM06}
\begin{align}
  \frac{\partial}{\partial\Lambda} U^\Lambda  
  & = \frac{(U^\Lambda)^2}{2 \pi}
  \sum_{\epsilon=\pm\Lambda} \left[ 
    \tilde G^\Lambda_\up(i \epsilon) \,
    \tilde G^\Lambda_\down(-i \epsilon) +
    \tilde G^\Lambda_\up(i \epsilon) \,
    \tilde G^\Lambda_\down(i \epsilon) \right] \nonumber \\ 
  & = \frac{2\, \left(U^\Lambda\right)^2\, 
    V_\up^\Lambda \, V_\down^\Lambda/\pi}
  {\left[ (\Lambda+ \Gamma)^2
      +(V_{\up}^\Lambda)^2 \right]
    \left[ (\Lambda+ \Gamma)^2
      +(V_{\down}^\Lambda)^2 \right]} \; ,
\end{align}
again with initial condition $U^{\Lambda=\infty}=U$.

A systematic improvement with respect to the previous results can be
observed in Fig.~\ref{fig:fig4}, with the dotted line resulting from
the inclusion of the interaction renormalization.  The quantitative
agreement with the exact results is excellent and holds for
$U/\Gamma=25$, the largest value with available Bethe ansatz
data \cite{Gerland00}.  Also for more complex dot geometries this
extension considerably improves the agreement with NRG
results \cite{KEM06}.  We note that a truncated fRG scheme including
the full frequency dependence of the two-particle vertex, and hence a
frequency dependent self-energy, reproduces also the Kondo resonance
and Hubbard bands in the spectral function \cite{HMPS04}. This
requires, however, a substantial computational effort.

\subsection{Effect of magnetic fields}

\begin{figure}[htbp]
  \centering
  \includegraphics[height=9.5cm,angle=-90,clip]{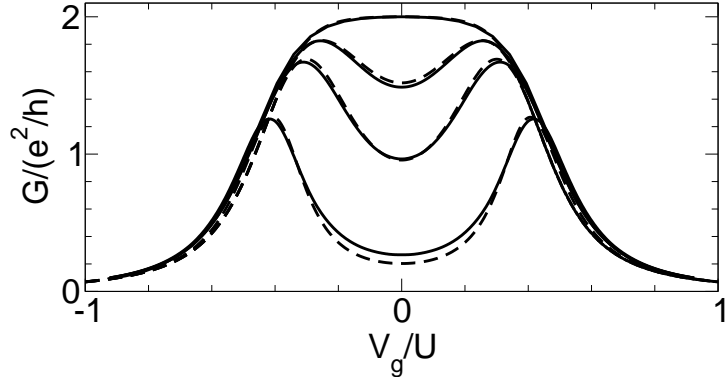}
  \caption{Gate voltage dependence of the total conductance $G$ of a single
    dot with $U/\Gamma=3\pi$ and $h/\Gamma=0$, $0.058$, $0.116$,
    $0.58$ from top to bottom. In units of the $V_g=0$ Kondo
    temperature $T_K^\NRG/\Gamma=0.116$ these fields correspond to
    $h=0$, $0.5 T_K^\NRG$, $T_K^\NRG$, and $5T_K^\NRG$.  Solid line:
    NRG data from \cite{Costi01}. Dashed lines: fRG
    approximation with flow of vertex.}
  \label{fig:fig5}
\end{figure}

\begin{figure}[htbp]
  \centering
  \includegraphics[height=9.5cm,angle=-90,clip]{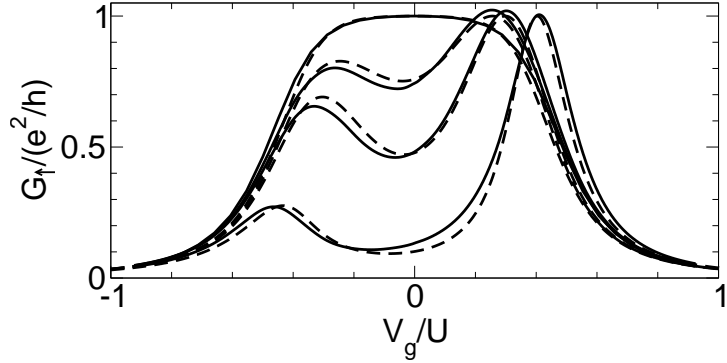}
  \caption{Gate voltage dependence of the partial conductance $G_\up$
    of a single dot for the same parameters as in Fig.~\ref{fig:fig5}.}
  \label{fig:fig6}
\end{figure}

We next consider the case of finite magnetic fields. For $h>0$ the
Kondo resonance in the NRG solution of the spectral function splits
into two peaks with a dip at $\omega=0$, resulting in a dip of
$G(V_g)$ at $V_g=0$. In Figs.~\ref{fig:fig5} and \ref{fig:fig6} we
compare the total $G=G_\up+G_\down$ and partial $G_\up$ conductance
obtained from the above fRG truncation scheme including the flow of
the effective interaction with NRG \cite{Costi01} results for different
$h$ expressed in units of $T_K^\NRG=0.116\Gamma$, where $T_K^\NRG$ is
the width of the Kondo resonance at the particle-hole symmetric point
$V_g=0$ obtained by NRG \cite{Costi01}.

The excellent agreement between NRG and fRG results provides strong
evidence of the presence of the Kondo scale within the truncated fRG
scheme.

\subsection{Determining the Kondo scale}

\begin{figure}[htbp]
  \centering
  \includegraphics[height=9.5cm,angle=-90,clip]{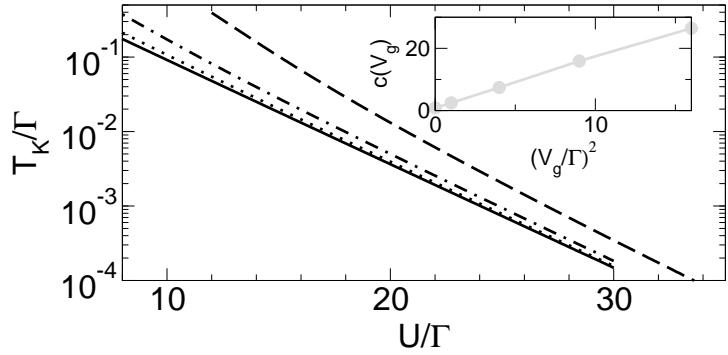}
  \caption{The Kondo scale $T_K$ as a function of $U$ for different $V_g$:
    $V_g=0$ (solid), $V_g/\Gamma=1$ (short dashed), $V_g/\Gamma=2$
    (dotted), and $V_g/\Gamma=4$ (long dashed). {\it Inset:} The
    fitting parameter $c$ as a function of $(V_g/\Gamma)^2$.}
  \label{fig:fig7}
\end{figure}

From the above comparison of fRG and NRG data shown in
Fig.~\ref{fig:fig6} for different $h$ we infer the appearance of an
exponentially small energy scale defining the Kondo scale $T_K(U,V_g)$
by the magnetic field required to suppress the total conductance to
one half of the unitary limit, $G(U,V_g,h=T_K)=e^2/h$.  For fixed
$U\gg \Gamma$ this definition applies for gate voltages within the
resonance plateau for $h=0$. In Fig.~\ref{fig:fig7} we show
$T_K(U,V_g)$ for different $V_g$ as a function of $U$ on a linear-log
scale.  The curves can fitted to a function of the form
\begin{align}
  \label{eq:fit}
  f(U/\Gamma) & = a \exp{\left[- \left| b \frac{U}{\Gamma} - c
        \frac{\Gamma}{U} \right| \right]} 
\end{align}
with $V_g$-dependent coefficients $a$, $b$, and $c$.  The above form
is consistent with the exact Kondo temperature $T_K$ \cite{Hewson} that
depends exponentially on a combination of $U$ and the level position
\begin{align}
  T_K^{\text{exact}} & \sim \exp{\left[- \frac{\pi}{2 U \Gamma} 
      \left|U^2/4-V_g^2\right|\right]}
  \notag \\
  \label{TKondo}
  & = \exp{\left[- \left| \frac{\pi}{8} \frac{U}{\Gamma} - 
        \frac{\pi}2 \frac{V_g^2}{\Gamma^2} \frac{\Gamma}{U} \right| \right] }
  \; .
\end{align}
The prefactor of the exponential depends on the details of the model
considered. To leading order its $U$ and $V_g$ dependence can be
neglected.  For the fit to Eq.~\eqref{eq:fit} we find $b(V_g)
\approx0.32$ for all $V_g$ (see Eq.~(\ref{eq:exp}) for comparison), in
good agreement with the exact value $\pi/8\approx 0.39$.  The
prefactor $a$ depends only weakly on $V_g$, and $c(V_g)$ increases
approximately quadratically with $V_g/\Gamma$ as shown in the inset of
Fig.~\ref{fig:fig7}, according to the behavior of the exact Kondo
temperature Eq.~\eqref{TKondo}. We thus conclude that $T_K(U,V_g)$ can
be determined from the $h$ dependence of the conductance obtained from
the fRG.  Similar results are obtained from the local spin
susceptibility \cite{Hewson}.

For the single dot at $T=0$ the exact conductance, transmission phase,
and dot occupancies are directly related by a generalized Friedel sum
rule \cite{Hewson}: $G_\sigma/(e^2/h)=\sin^2{(\pi \left< n_\sigma
  \right> )}$, and the transmission phase $\alpha_\sigma=\pi \left<
  n_\sigma \right>$.  As $0 \leq \left< n_\sigma \right> \leq 1$ the
argument of $\sin^2$ is restricted to a single period and the relation
between $G_\sigma$, $\left< n_\sigma \right>$, and $\alpha_\sigma$ is
unique.  In many approximation schemes the Friedel sum rule does not
hold exactly.  In contrast, within our method we map the many-body
problem onto an effective single-particle one for which the Friedel
sum rule is fulfilled.  For gate voltages within the $h=0$ conductance
plateau the (spin independent) dot filling is $1/2$ and the (spin
independent) phase is $\pi/2$. For sufficiently large $U/\Gamma$ the
crossover to $\left< n_\sigma \right>=1$ and $\alpha_\sigma=\pi$ to
the left of the plateau as well as $\left< n_\sigma \right>=0$ and
$\alpha_\sigma=0$ to the right is fairly sharp.


\section{Summary and Conclusion}
\label{sec:outlook}

We presented an fRG scheme developed for the study of electronic
transport through quantum-dot systems.  The resulting differential
flow equations describe the effective renormalized level position and
interaction in presence of local Coulomb interactions and magnetic
fields.  Analytical estimates capture signatures of the Kondo effect,
and a comparison with exact Bethe ansatz and high-precision NRG
results shows excellent agreement up to the largest Coulomb
interaction for which Bethe ansatz or NRG data are available.

The presented fRG scheme constitutes a reliable and promising tool in
the investigation of correlation effects and interference phenomena in
quantum-dot systems. The application to a single quantum dot can be
directly extended to different geometries involving more complicated
setups \cite{KEM06} such as parallel double dots \cite{MM06}.  In
\cite{KHODM06} this method was used to investigate the
long-standing phase lapse puzzle of the transmission phase through
multi-level quantum dots.

Furthermore, the truncated fRG scheme was sucessfully used to study
one-dimensional fermionic lattice models with two-particle interaction
and impurities \cite{MMSS02a,AEMMSS04,EMABMS04,AEMMSS06} (inhomogeneous
Luttinger liquids). In addition to the universal low-energy
asymptotics the fRG captures nonuniversal properties at higher energy
scales.  Novel low-energy fixed points were found for Y-junctions of
one-dimensional quantum wires pierced by a magnetic flux \cite{BSMS04}.
Also a single dot with Luttinger liquid leads \cite{AEM06} was
investigated. For the latter the competition between Kondo and
Luttinger liquid physics leads to a broad resonance plateau for all
experimentally accessible length scales, whereas the low-energy fixed
point is described by a sharp resonance characteristic for the
Luttinger liquid behavior.


\begin{acknowledgement}
  We thank X. Barnab\'e-Th\'eriault, R. Hedden, W. Metzner, Th.
  Pruschke, H. Schoeller, U. Schollw\"ock, and K. Sch\"onhammer for
  useful discussions.  We thank T. Costi and J. von Delft for
  providing their NRG and Bethe ansatz data.  This work has been
  supported by the French ANR (program PNANO) (S.A.), the Deutsche
  Forschungsgemeinschaft (SFB 602) (V.M.), and a Feoder Lynen
  fellowship of the Alexander von Humboldt foundation and the Istituto
  Nazionale di Fisica della Materia--SMC--CNR (T.E.).
\end{acknowledgement}


\end{document}